# Hadron electromagnetic structure: Shedding light on models and their mechanisms

Derek B. Leinweber[a]

[a]Department of Physics, The Ohio State University, 174 West 18th Avenue, Columbus, OH 43210–1106

Strange quark contributions to the proton magnetic moment are estimated from a consideration of baryon magnetic moment sum rules. The environment sensitivity of quark contributions to baryon moments is emphasized. Pion cloud contributions to proton charge radii are examined in the framework of Chiral Perturbation Theory. The absence of scalar-diquark clustering in the nucleon is discussed.

## 1. INTRODUCTION

One of the great promises of the lattice gauge approach to QCD is to reveal the quark substructure and dynamics of hadrons. By examining the manner in which quarks contribute to various observables one can identify which "QCD-inspired" models truly reflect the properties of QCD, and discard those models which do not. By identifying models which reproduce QCD properties, one can discover the underlying mechanisms which give rise to the quark dynamics revealed in the lattice calculations.

Here I will review three aspects of the electromagnetic structure of QCD on which the lattice calculations have shed some light. Quark contributions to baryon magnetic moments are discussed with an emphasis on environment sensitivity. An estimate of strange quark contributions to the proton moment is obtained through the consideration of magnetic moment sum rules. Pion cloud contributions to hadron charge radii are discussed, and the results of a search for diquark clustering in the proton are presented.

## 2. MAGNETIC MOMENT SUM RULES

While the simple quark model has been qualitatively successful in predicting the magnetic moments of octet baryons, the model requires significant modifications under closer examination. For example, the following sum rule for baryon magnetic moments follows as a consequence of the predictions of the SU(6) spin-flavor symmetric quark model broken only by the quark masses

$$\frac{3(p - \Sigma^+)}{\Xi^- - \Xi^0} = \frac{p + 3\Lambda}{p} \, . \quad (1)$$

In fact, the ratio of left/right-hand sides of this sum rule is $5.5 \pm 0.4$ for experimental moments as opposed to 1. The lattice results of Ref. [1] yield a ratio of $4.1 \pm 1.5$.

A detailed structure in the quark contributions to octet-baryon moments has been revealed by the lattice calculations of [1]. Figure 1 summarizes effective quark moments defined by equating lattice quark sector contributions and SU(6) predictions. This is not a nonrelativistic approximation, but rather a way of presenting the lattice results in a familiar context. In addition to reproducing the observed baryon moments, a successful model of hadron structure will reproduce the environment sensitivity of the quark moments illustrated in Figure 1. For example, the $u$-quark moment is much smaller in the environment of the two strange quarks of $\Xi$ than in the proton.

While there are many magnetic moment sum rules whose experimental violations are reproduced by the current lattice QCD predictions, the lattice violation of the Sachs sum rule,

$$3(p + n) = \Sigma^+ - \Sigma^- + \Xi^0 - \Xi^- \, , \quad (2)$$

is not in accord with experiment [2]. The lattice moments yield a ratio of left/right-hand sides of 1.29(20), whereas the experimental moments indicate the ratio is 0.881(11).

In the lattice calculations of Ref. [1] the contributions from diagrams in which the electromag-netic current first interacts with a disconnected quark loop are not included. These loop contributions, $\mu_l$, are equal among octet baryons of a given strangeness when the $u$ and $d$ quarks are taken with the same mass. The experimental violation of (2) indicates

$$\frac{3(p+n)+6\mu_l}{\Sigma^+ - \Sigma^- + \Xi^0 - \Xi^-} = 0.881(11), \qquad (3)$$

and the lattice results yield $\mu_l = -0.19(9)\ \mu_N$. Assuming a baryon independence of the loop contributions, a fit to the baryon octet suggests $\mu_l = -0.10(6)\ \mu_N$.

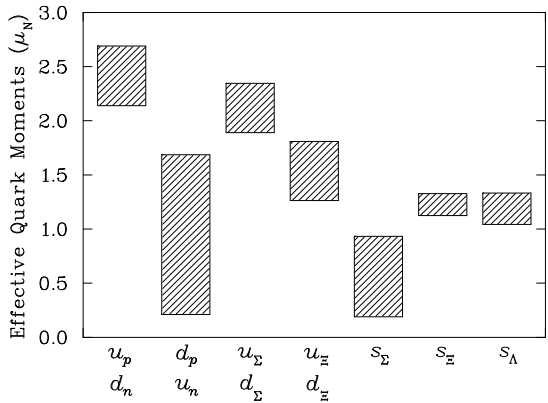

Figure 1. Effective moments normalized to unit charge for quarks residing in octet baryons. The statistical uncertainties reflect the sensitivity of the moments to nonperturbative gluon dynamics.

The relative role of strange and light quarks in the disconnected loop may be estimated through the use of hadronic models [3]. There the contribution of the heavier strange quark is suppressed by a factor of two relative to the $d$ quark. This places the lattice estimate of the strange quark contribution to the proton magnetic moment at $\mu_s \simeq +0.1$ to $+0.2\ \mu_N$. Ab initio calculations of this quantity are anticipated in the near future.

## 3. THE PION CLOUD

The role of the pion cloud in reproducing hadron properties is a central issue of hadron phenomenology. Charge radii are particularly interesting due to the weight of long distance physics in this observable.

Lattice calculations of charge radii have historically produced values for the pion and proton which are roughly equal at 0.65(8) fm. While the pion result agrees with the experimental radius of 0.66(1), the lattice proton radius is small relative to 0.862(12) fm. A possible solution to this long-standing problem lies in chiral corrections to the extrapolation of charge radii to the physical quark masses [4].

Conventional Chiral Perturbation Theory suggests the following analytic structure for pion and proton radii

$$\begin{aligned}
\langle r_\pi^2 \rangle &= \frac{1}{(4\pi)^2} \frac{1}{f_\pi^2} \ln\left(\frac{1}{m_\pi^2 a^2}\right) + c_0^\pi + c_2^\pi m_\pi^2 \\
\langle r_p^2 \rangle &= \frac{1}{2}\left(\frac{g_{\pi NN}^2}{4\pi}\frac{3}{2\pi M^2} + \frac{1}{(4\pi)^2}\frac{1-g_A^2}{f_\pi^2}\right) \\
&\quad \times \ln\left(\frac{1}{m_\pi^2 a^2}\right) + c_0^p + c_2^p m_\pi^2 \qquad (4)
\end{aligned}$$

Taking physical values for the parameters and fitting $c_0$ and $c_2$ to the lattice results allows one to estimate the importance of the pion cloud. While the corrections are small for the pion, they are large for the proton. Figure 2 illustrates the conventional linear and standard chiral extrapolations. The pion cloud makes significant contributions to the proton charge radius. Any realistic model of the proton will necessarily have to incorporate this long distance physics.

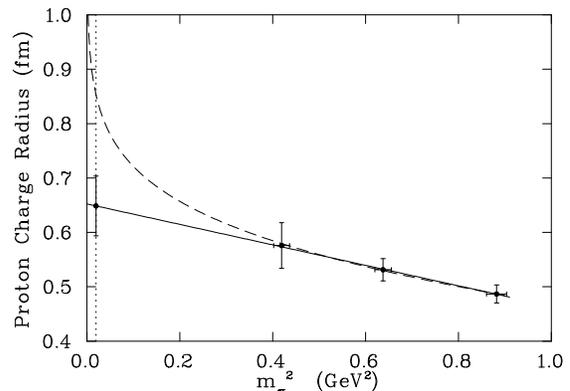

Figure 2. The conventional linear and standard chiral extrapolations of the proton charge radius.

## 4. DIQUARK CLUSTER SEARCH

The early success of hyperfine interactions based on single-gluon exchange, has led some authors to speculate that the strong attractive hyperfine interaction pairs a $u$ and $d$ quark of the nucleon into a scalar diquark of considerable clustering. It is often argued that point-like scalar diquarks naturally arise in QCD.

A gauge invariant method for the examination of scalar diquark clustering in the nucleon ground state has been identified [5]. Direct evidence indicating the absence of substantial diquark clustering of $u$ and $d$ quarks in low-lying baryons is obtained by examining the quark distributions in octet baryons [1], where quarks may form scalar diquarks, and comparing these distributions with the relevant decuplet baryons [6], where quarks are predominantly paired in vector diquarks.

Consider the quark distributions within the proton and how they will change in going from $p$ to $\Delta^+$. In the quark-diquark model the scalar diquark cluster is lost in $\Delta^+$. The net positive charge of the $u$-$d$ cluster in the proton moves to larger radii, and increases the $\Delta^+$ radius. Both $u$- and $d$-quark charge distributions are predicted to swell in going to $\Delta^+$. Unfortunately, a comparison of $p$ and $\Delta^+$ charge radii in a diquark model does not appear to have been considered. To estimate a lower bound for the size of these anticipated quark distribution swellings, we refer to the nonrelativistic quark model where the role of hyperfine attraction plays a much weaker and less dramatic role.

In the nonrelativistic quark model, both the $u$- and $d$-quark distributions become broader as the attraction between $u$ and $d$ quarks is replaced by hyperfine repulsion in $\Delta$. In the model of Isgur-Karl-Koniuk [7] the predicted increase in the rms charge radius is [8] $r_\Delta/r_p = 1.28$, with the quark distributions experiencing a large swelling of $r_\Delta^u/r_p^u = 1.33$ and $r_\Delta^d/r_p^d = 1.49$.

The physics of light pions discussed in the previous section is not included in either of the quark models considered here. To allow a comparison with these models on a more equal footing, the lattice extrapolations of charge radii are done linearly in $m_\pi^2$ from where $\Delta$ is stable, effectively subtracting the nonanalytic contributions associated with light pion dressings from the charge radii.

The lattice results suggest that the $\Delta^+$ charge radius may actually be *smaller* than that of the proton, $r_\Delta/r_p = 0.97(7)$. This result differs by four standard deviations from the prediction of the nonrelativistic quark model. The quark distribution radii indicate the dominant effect in the lattice results is the broadening of the negatively charged $d$-quark distribution, as $r_\Delta^u/r_p^u = 1.01(9)$ and $r_\Delta^d/r_p^d = 1.12(16)$. Additional details and discussion may be found in Ref. [5].

The striking difference between the lattice results and the two models considered here is *the absence of any significant change in the lattice $u$-quark distribution radius*. In both models, the $u$-quark distribution was predicted to be broader in $\Delta$, largely due to the disappearance of scalar diquark clustering in going from the nucleon to $\Delta$. The lattice results indicate that hyperfine attraction does not lead to substantial scalar diquark clustering in the nucleon ground state. Diquark cluster models are not supported as an appropriate description of the internal structure of low-lying baryons.

This research is supported by the Department of Energy and the National Science Foundation.